\documentclass[reprint,aps,prb,amsmath,amssymb]{revtex4-1}

\usepackage[]{graphicx}
\usepackage[]{units}
\usepackage{times}
\usepackage{bm}
\usepackage{ulem}		
\usepackage{color}
\renewcommand{\bm }{\mathbf}

\newcommand{\comment}[1]{}

\renewcommand{\emph}{\textit}

\begin{document}

\title{Dark and bright exciton formation, thermalization, and photoluminescence in monolayer transition metal dichalcogenides}
\author{Malte Selig$^{1,2}$}
\author{Gunnar Bergh\"auser$^2$}
\author{Marten Richter$^1$}
\author{Rudolf Bratschitsch$^3$}
\author{Andreas Knorr$^1$}
\author{Ermin Malic$^2$}
\affiliation{$^1$Institut f\"ur Theoretische Physik, Technische Universit\"at Berlin,  10623 Berlin, Germany}
\affiliation{$^2$Chalmers University of Technology, Department of Physics, SE-412 96 Gothenburg, Sweden}
\affiliation{$^3$Institute of Physics and Center for Nanotechnology, University of M\"unster, 48149 M\"unster, Germany}

\begin{abstract}
The remarkably strong Coulomb interaction in atomically thin transition metal dichalcogenides (TMDs) results in an extraordinarily rich many-particle physics including the formation of tightly bound excitons. Besides  optically accessible bright excitonic states, these materials also exhibit a variety of dark excitons. Since they can even lie below the bright states, they have a strong influence on the exciton dynamics, lifetimes, and photoluminescence. 
While very recently, the presence of dark excitonic states has been experimentally demonstrated, the origin of these states, their formation, and dynamics have not been revealed yet.  Here, we present a microscopic study shedding light on time- and energy-resolved formation and thermalization of bright and dark intra- and intervalley excitons as well as their impact on the photoluminescence in different TMD materials. We demonstrate that intervalley dark excitons, so far widely overlooked in current literature, play a crucial role in 
tungsten-based TMDs giving rise to an enhanced photoluminescence and reduced exciton lifetimes at elevated temperatures.

\end{abstract}

\maketitle


Monolayers of transition metal dichalcogenides (TMDs) have attracted much attention in current research due to their remarkable optical and electronic properties making them interesting nanomaterials for both fundamental science and technological applications.\cite{Mak2010,Cao2012,Butler2013,He2014,Schmidt2016,Neumann2017} As atomically thin materials, they show a reduced screening of the Coulomb potential resulting in an extraordinarily strong Coulomb interaction and the formation of tightly bound excitons.\cite{Steinhoff2014,Ramasub2012,Chernikov2014,Berghauser2014} Furthermore, they exhibit an efficient light-matter interaction giving rise to pronounced features in optical spectra.\cite{Mak2010,He2014,AroraWSe22015,AroraMoSe22015} Very recently, the presence of energetically dark exciton states has been experimentally demonstrated.\cite{Zhang2015,AroraWSe22015,AroraMoSe22015,Zhang2016,Molas2016} So far, only one class of such dark states has been considered in literature, namely spin-forbidden states, where the Coulomb-bound electron and hole states have opposite spin.\cite{Kormanyos2015, Plechinger2016} However, the formation of excitons is not limited to the same high-symmetry point in the Brillouin zone allowing for a different class of dark states with a non-zero center-of-mass momentum.\cite{Wu2015,Qiu2015,Selig2016}
Since these momentum-forbidden intervalley dark excitonic states can lie below the optical bright excitons,\cite{Selig2016} exciton-phonon scattering into these states is highly efficient even at very low temperatures and determines the excitonic dephasing time and the homogeneous linewidth of these materials.\cite{Selig2016} In contrast to the previously studied spin-forbidden dark states, they do not require a spin-flip and thus, the exciton-phonon scattering occurs on a femtosecond time scale.

\begin{figure}[t!]
 \begin{center}
\includegraphics[width=\linewidth]{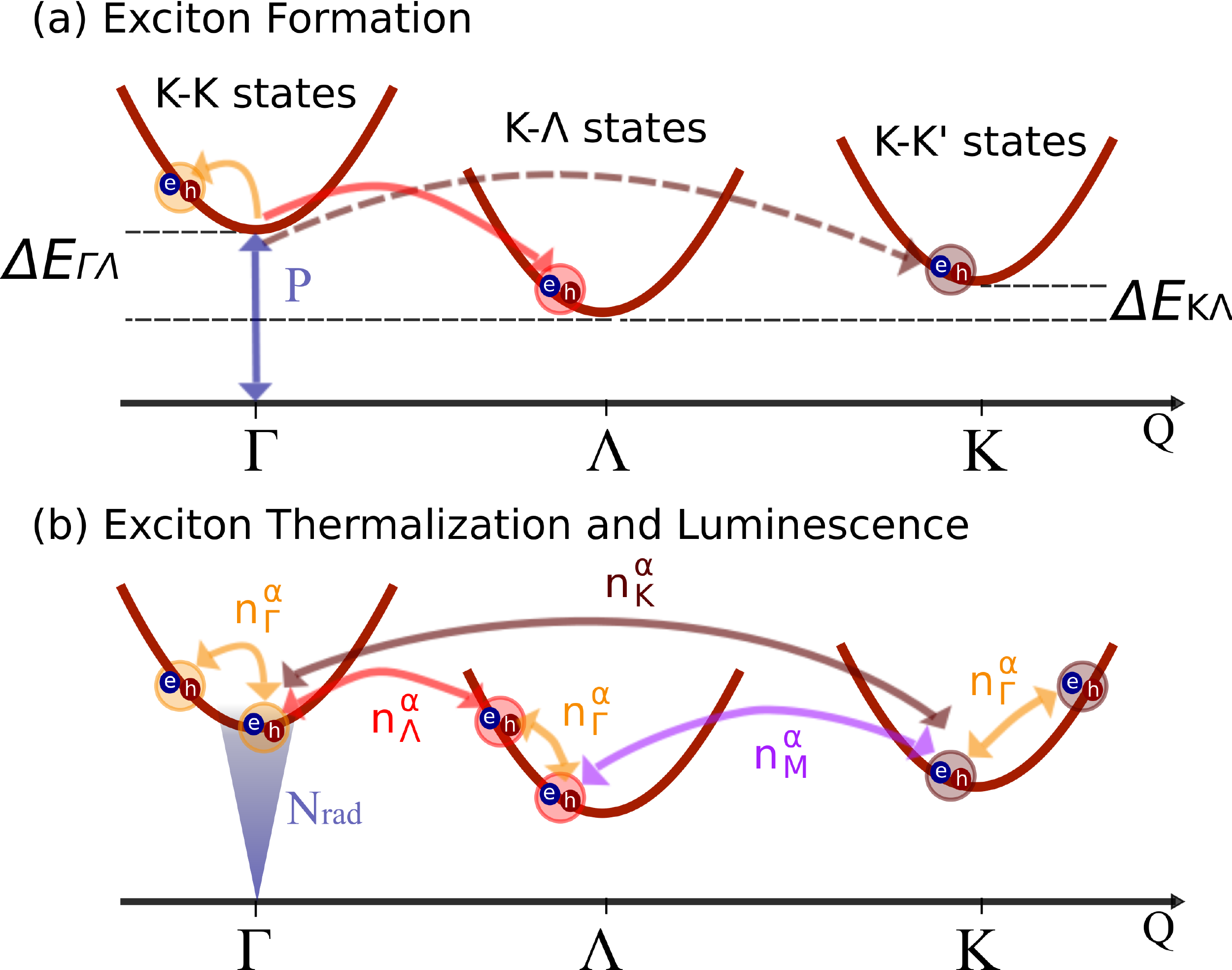}
 \end{center}
 \caption{\textbf{Exciton formation, thermalization, and luminescence.} (a) After optically exciting a coherent exciton population $P$, incoherent excitons are formed in different valleys assisted by emission and absorption of phonons. We take into account intravalley excitons, where the Coulomb bound electron and hole are both placed at the $K$ valley as well as intervalley excitons, where the hole lies at the $K$ point, while the electron is either at the $\Lambda$, or $K'$ valley. The corresponding states denoted as K-K, K-$\Lambda$, and K-K' excitons are located at the $\Gamma$, $\Lambda$,  and $K$ point in the excitonic Brillouin zone that is characterized by the center-of-mass momentum $\mathbf{Q}$.
(b) Intra- and intervalley exciton-phonon scattering leads to thermalization of incoherent exciton populations. Excitons located within the light cone (blue shaded) can decay via radiative recombination $N_{\text{rad}}$ resulting in photoluminescence. $n^{\alpha}_i$ denotes the occupation of the involved phonon from the $i=\Gamma,\Lambda,K,M$ point in the 1. BZ..}
 \label{schema}
\end{figure}

Here, we present a sophisticated many-particle study of the time- and energy-resolved exciton dynamics in monolayer TMDs. In particular, we shed light on the phonon-assisted formation and thermalization of excitons as well as on their impact on the photoluminescence yield. Besides the bright excitons, where the Coulomb bound electron and hole are both placed at the $K$ valley, we also take into account dark excitonic states with a center-of-mass momentum exceeding the light cone. This includes intervalley dark excitons, where the hole lies at the $K$ valley, while the electron is located either at the $\Lambda$, $\Lambda'$, or $K'$ valley, cf. Fig. \ref{schema}. 
 We find that first on a timescale of \unit[100]{fs} a coherent exciton population is generated within the light cone decaying via radiative recombination and exciton-phonon scattering. Then  on a sub-picosecond timescale, incoherent exciton populations in K-K, K-$\Lambda$ and K-K' states are generated via emission and absorption of acoustic and optical phonons. After thermalization, most excitons are located in the energetically lowest K-$\Lambda$ (K-K) states in tungsten(molybdenum)-based TMDs. We show that this has a crucial impact on the photoluminescence resulting in an increasing (decreasing) luminescence yield at elevated temperatures for WSe$_2$ (MoSe$_2$) -- in contrast to the behavior of conventional semiconductors, such as GaAs quantum wells, and in excellent agreement with very recent experimental observations.\cite{,AroraWSe22015,AroraMoSe22015,Zhang2015}

The first step towards modeling the exciton dynamics in TMDs is the solution of the  Wannier equation\cite{Berghauser2014,Kira2006,Selig2016,Kochbuch,Thranhardt2000,Knorr1996}
\begin{equation}
\frac{\hbar^2 \mathbf{q}^2}{2 m} \varphi^{\mu}_\mathbf{q} - \sum_\mathbf{k} V^{\text{exc}}_\mathbf{q,k} \varphi^{\mu}_\mathbf{q+k}=(E^{\mu}-E_0) \varphi^{\mu}_\mathbf{q}\label{Wannier}
\end{equation}
providing access to all excitonic states $\mu$ including their energies $E^{\mu}$ and wave functions $\varphi^{\mu}_\mathbf{q}$. Here, $V^{\text{exc}}_\mathbf{q,k}$ denotes the attractive part of the Coulomb interaction giving rise to the formation of excitons,  while $E_{0}$ stands for the gap energy of the Coulomb-renormalized electronic states, and $m$ for the reduced excitonic mass. The solution of the Wannier equation explicitly includes besides the bright excitons also dark intra- and intervalley states depicted in Fig. \ref{schema}. We do not consider excitons with holes located at the $K'$ point, since besides the different light polarization these states are symmetric to already considered K-K' excitons. Furthermore, spin-forbidden dark states requiring spin-flip processes are beyond the scope of this study and will be in the focus of future work.

 The next step is to derive equations of motion for  the excitonic polarization $P_\mathbf{Q}^{\mu}(t)=\sum_\mathbf{q} \varphi^{* \mu}_\mathbf{q} \langle a^{\dagger v}_\mathbf{q + \beta Q} a^{c}_\mathbf{q-\alpha Q} \rangle(t)$ \cite{Kira2006,Thranhardt2000} being a measure for optically induced interband transitions and  the incoherent exciton occupation $ N_\mathbf{Q}^{\mu}(t)=\sum_\mathbf{q,q'} \varphi^{* \mu}_\mathbf{q'} \varphi^{\mu}_\mathbf{q} \delta \langle a^{\dagger c}_\mathbf{q-\alpha Q} a^{v}_\mathbf{q+\beta Q} a^{\dagger v}_\mathbf{q'+\beta Q} a^{c}_\mathbf{q'-\alpha Q} \rangle(t)$ corresponding to the expectation value of a correlated electron-hole pair\cite{Thranhardt2000}. Since we consider only terms up to the second order in the exciting electromagnetic field, we neglect the electron (hole) densities in this study. Here $a^{(\dagger)\lambda}_\mathbf{k}$ denotes the annihilation (creation) operators with momentum $\mathbf{k}$ and band $\lambda=c,v$. Furthermore, we have introduced relative $\bm q$ and center-of-mass momenta $\bm Q$ with the coefficients $\alpha = m_\text{e}/(m_\text{h}+m_\text{e})$ and $\beta = m_\text{h}/(m_\text{h}+m_\text{e})$ describing the relative electron and hole masses. 
While the temporal evolution of the excitonic polarization $P_\mathbf{Q}^{\mu}(t)$ is driven by an external optical field giving rise to a coherent exciton occupation,  the  dynamic of $ N_\mathbf{Q}^{\mu}(t)$  describes the formation and thermalization of  incoherent excitons induced by a non-radiative decay of  coherent excitonic states.
The corresponding time- and momentum-dependent equations for $P_\mathbf{Q}^{\mu}(t)$ and 
$ N_\mathbf{Q}^{\mu}(t)$ can be found in the appendix (Eqs. (3) and (4), respectively).

\begin{figure}[t!]
 \begin{center}
\includegraphics[width=\linewidth]{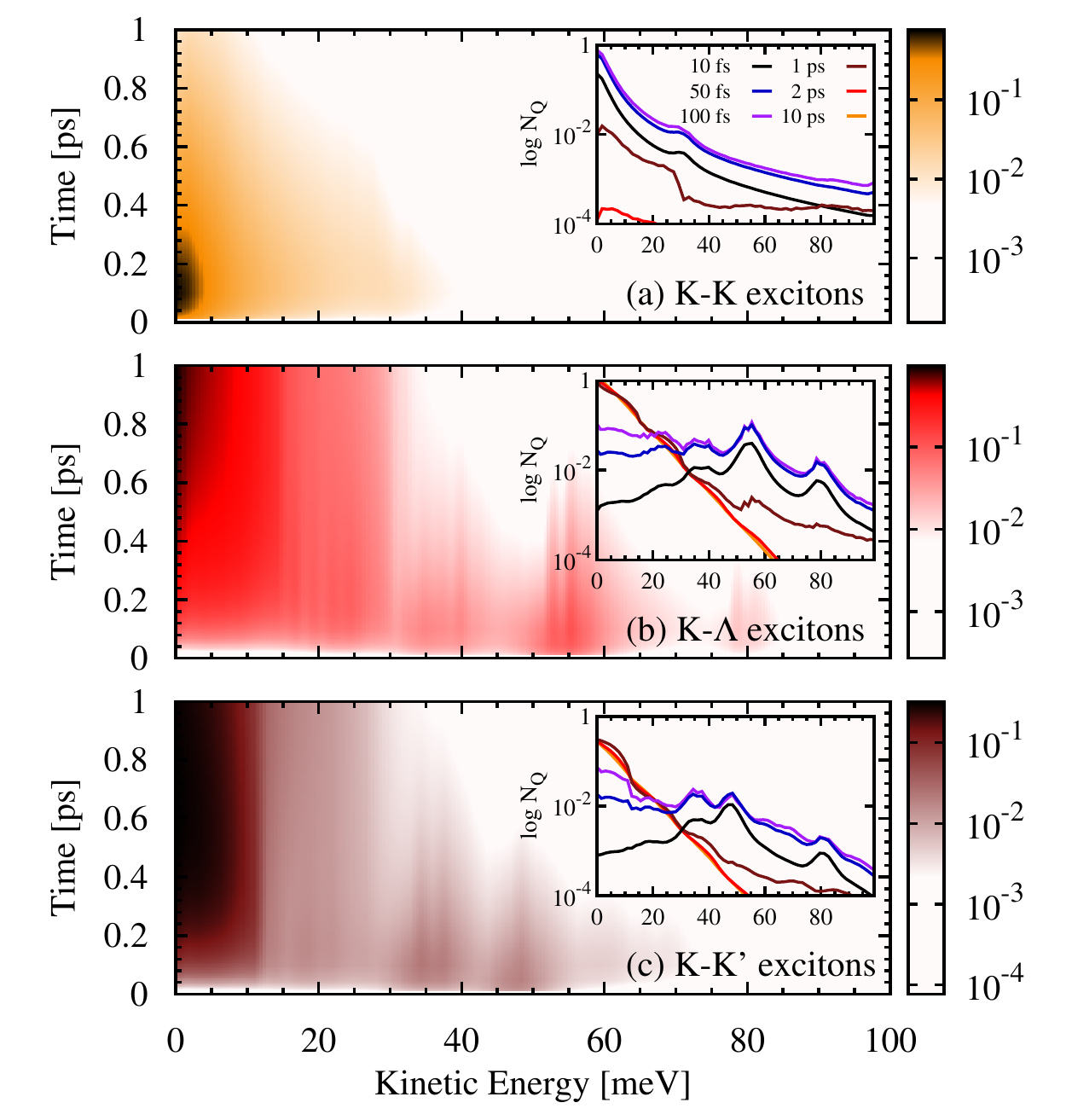}
 \end{center}
 \caption{\textbf{Exciton dynamics in WSe$_2$.} Time-dependent incoherent exciton occupation $ N^{\mu}_\mathbf{Q}$ as a function of the excitonic kinetic energy (energies are shown with respect to the corresponding valley minimum) at the exemplary temperature of \unit[77]{K} for the (a) K-K, (b) K-$\Lambda$, and (c) K-K' excitonic states. Logarithmic time snapshots are shown in the insets and illustrate the exciton formation (increase of $ N^{\mu}_\mathbf{Q}$) up to approximately \unit[100]{fs} followed by exciton thermalization on a picosecond timescale. }
 \label{excDyn}
\end{figure}

To obtain access to the photoluminescence of TMDs, we evaluate the quantum mechanical expression for the intensity of the emitted light\cite{Thranhardt2000} $I(t)\propto\sum_{\mathbf{K},\sigma} \partial_t n^{\sigma}_\mathbf{K}(t)$ with the photon density
$
n_\mathbf{k}^{\sigma}(t)=\langle c_\mathbf{k}^{\dagger \sigma} c_\mathbf{k}^{\sigma} \rangle(t)$, where
 $c_\mathbf{k}^{(\dagger) \sigma}$ are the photon annihilation (creation) operators with the photon momentum $\mathbf{k}$ and polarization $\sigma$. 
The equation of motion for the emitted intensity reads
\begin{align}
I(t)=\frac{2 \pi}{\hbar}\hspace{-2pt}\sum_{\mathbf{k},\sigma} |M^{\sigma}_{\mathbf{Q},\mathbf{k}}|^2 \left(  |P_\mathbf{Q}(t)|^2 +  N_\mathbf{Q}(t)\right)\delta( \Delta E^\sigma_\mathbf{k} )\delta_\mathbf{Q,k_\parallel}.\label{eqPL}
\end{align}
Here, the sum runs over all three-dimensional photon momenta $\mathbf{k}$ with the restriction that the absolute value is  smaller than the light cone radius $K_c$. Furthermore, the appearing Dirac and Kronecker delta make sure that only excitons fulfilling both the momentum conservation (i.e. the transverse photon momentum $\mathbf{k}_\parallel$ coincides with the exciton momentum $\mathbf{Q}$) and energy conservation $ \Delta E^\sigma_\mathbf{k}=(E_\mathbf{k_\parallel} -\hbar \omega_\mathbf{k}^{\sigma})$ can contribute to the emission. The first term in Eq. (\ref{eqPL}) is driven by the coherent exciton occupations $|P_\mathbf{Q}|^2$, while the second term is determined by the incoherent exciton densities $ N_\mathbf{Q}$. 
As a result, the numerical solution of the coupled set of differential equations for  the  excitonic polarization $P_\mathbf{Q}^{\mu}(t)$ and  the incoherent exciton occupation $ N_\mathbf{Q}^{\mu}(t)$
allows us to study the formation and thermalization of bright and dark intra- and intervalley excitons as well as their impact on the photoluminescence.

After resonant optical excitation with a delta-shaped pulse, a coherent exciton occupation $|P_\mathbf{0}|^2$ at the $\Gamma$ valley is formed with an approximately zero center-of-mass momentum $\bm Q \approx 0$. The occupation decays due to radiative recombination and exciton-phonon scattering within the excitonic $\Gamma$ valley as well as into dark excitonic K-$\Lambda$ and K-K' states (Fig. \ref{schema}) resulting in a decay time in the range of tens of femtoseconds at room temperature.\cite{Selig2016} Exciton-phonon scattering also accounts for the formation of incoherent exciton occupations $ N_\mathbf{Q}^{\mu}$. 
Note that the total number of coherent and incoherent excitons $|P_\mathbf{0}|^2+ \sum_{\mathbf{Q},\mu}  N^\mu_\mathbf{Q}$ is a conserved quantity provided that the radiative recombination is negligibly small.

Figure \ref{excDyn} illustrates the time- and energy-dependent dynamics of the incoherent exciton occupations $ N^{\mu}_\mathbf{Q}(t)$ after an optical excitation in WSe$_2$ at an exemplary temperature of \unit[77]{K} in vicinity of the most relevant high-symmetry points in the excitonic Brillouin zone. 
For times up to approximately \unit[100]{fs} we observe the formation of incoherent exciton occupations due to non-radiative decay of the coherent exciton population $|P_\mathbf{0}|^2$, which is centered at $\bm Q \approx 0$ within the light cone (Fig. \ref {schema}). At the beginning, the formation of incoherent excitons occurs mainly in the K-K states due to emission and absorption of $\Gamma$ acoustic and optical phonons (Fig. \ref{excDyn} (a)) and in $K-\Lambda$ states due to the scattering with large momentum $\Lambda$ acoustic and optical phonons (Fig. \ref{excDyn} (b)). 
There is also a weaker initial formation of  K-K' excitons due to scattering with large-momentum $K$ phonons (Fig. \ref{excDyn} (c)). 
Note however that while incoherent K-K excitons are strongly centered at small kinetic energies, K-$\Lambda$ and K-K' excitons show an increased occupation especially at higher energies in the first \unit[100]{fs}. Here,  acoustic and optical phonons with a larger momentum are required to reach the dark intervalley excitons. The latter are located approximately $\Delta E_{K \Lambda}=$\unit[67]{meV} and $\Delta E_{K K'}=$\unit[64]{meV} below the bright state (Fig. \ref{schema} (a)) according to the solution of the Wannier equation, cf. Eq. (\ref{Wannier}). 

A closer look at the formation of K-$\Lambda$ excitons reveals two energetic regions with high occupations (cf. the inset of Fig. \ref{excDyn}(b)): First,  \unit[40]{meV} above the minimal exciton energy corresponding to the emission of optical $\Lambda$TO and $\Lambda$LO phonons,\cite{Li2013,Jin2014} which have energies about \unit[30]{meV}. Second, \unit[55]{meV} above the minimum corresponding to the emission of acoustic $\Lambda$LA and $\Lambda$TA phonons,\cite{Li2013,Jin2014} with energies around \unit[12]{meV}. Furthermore, we also find a small impact of the absorption of acoustic phonons at \unit[80]{meV}. Since phonon absorption is proportional to the Bose distribution $n_\mathbf{q}^{\alpha}$ with the phonon mode $\alpha$ and momentum $\bf q$, this feature is less pronounced than the emission peaks that are proportional to $(1+n_\mathbf{q}^{\alpha})$. Note that the energetic position of these peaks does not exactly correspond to the phonon energy, since exciton thermalization sets in even before the exciton formation has been finalized. As a result, excitons relax to lower kinetic energies and scatter from there with phonons. The features characterizing the spectral distribution of the K-K' exciton occupation can be explained in an analogous way.

After the characteristic phonon-induced formation time of approximately \unit[100]{fs}, thermalization of incoherent exciton occupations takes over and determines the dynamics of excitons. 
The generated hot K-K, K-$\Lambda$, and K-K' excitons scatter towards the energetically lowest states. This process is again driven by phonon-assisted scattering to dark intra- and intervalley excitonic states.  The steady-state exciton occupations correspond to degenerate Bose-Einstein distributions.

\begin{figure}[t!]
 \begin{center}
\includegraphics[width=\linewidth]{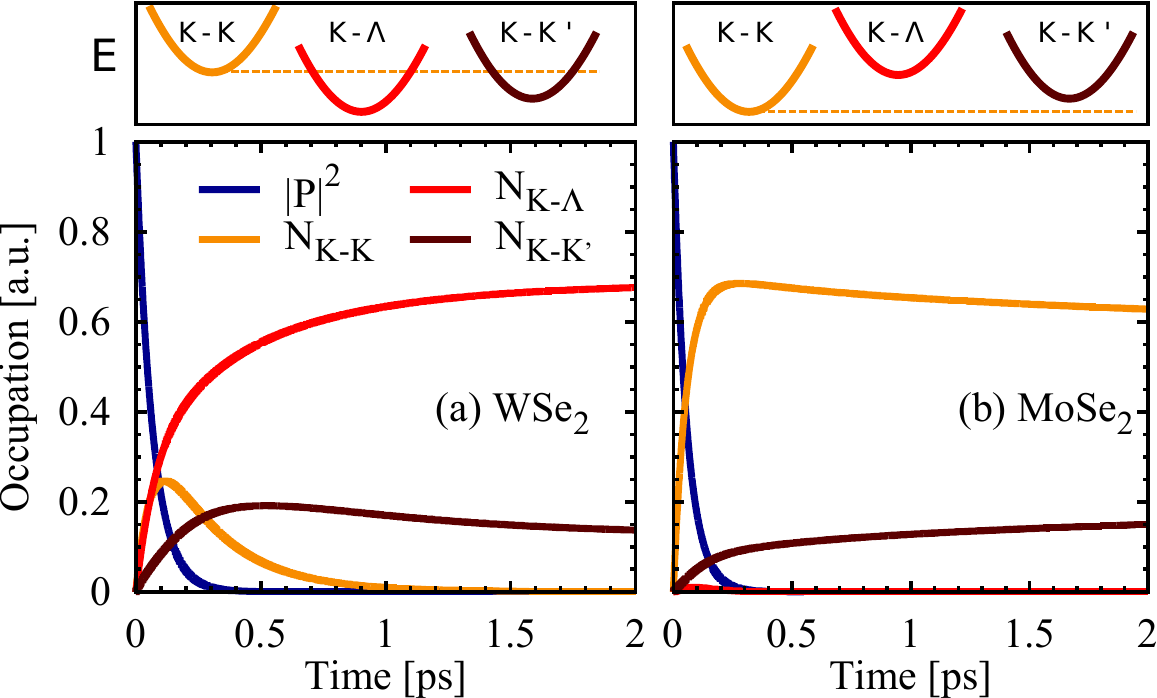}
 \end{center}
 \caption{\textbf{Exciton densities.} Temporal evolution of coherent $|P_\mathbf{0}|^2$ and incoherent exciton densities $ N_\mu=\sum_{\mathbf{Q}} N^\mu_\mathbf{Q}/A$ with $\mu=$K-K, K-$\Lambda$, and K-K' in (a) WSe$_2$ and (b) MoSe$_2$ at \unit[77]{K}. The largest densities are found in the energetically lowest excitonic states, i.e. K-$\Lambda$  and  K-K excitons in WSe$_2$ and MoSe$_2$, respectively. The upper panels illustrate the relative energetic positions of the different exciton states.}
 \label{valleys}
\end{figure}

To further study the thermalization dynamics, we determine the  temporal evolution of exciton densities  $ N_\mu=\sum_{\mathbf{Q}} N^{\mu}_\mathbf{Q}$ for different excitonic states $\mu=$ K-K, K-$\Lambda$, K-K' including the coherent exciton density $|P_\mathbf{0}|^2$, cf. Fig. \ref{valleys}(a). Here, the center-of-mass momentum is restricted to the corresponding valley in the excitonic Brillouin zone, i.e. $\mathbf{Q}\in (\Gamma, \Lambda, \text{or}\, K)$.
We reveal a thermalization time of approximately \unit[2]{ps}, after which the incoherent exciton occupations remain constant. Remarkably, the reached steady state occupation of K-K excitons is negligibly small. In contrast, the K-$\Lambda$ excitons show the largest density followed by K-K' excitons, where $ N_{K-\Lambda}$ is more than three times larger than $ N_{K-K'}$.  At first sight, this is surprising, since the energy separation between the K-K' and the K-$\Lambda$ excitons is only about $\Delta E_{K \Lambda}=\unit[3]{meV}$ (Fig. \ref{schema} (a)). However, it can be traced back to the fact that the $\Lambda$ point occurs three times in the Brillouin zone, while there is only one $K'$ point. Furthermore, the K-K' exciton occupation is interestingly not mainly driven by the decay of the coherent exciton occupation $|P_\mathbf{0}|^2$, but is rather formed indirectly through the exciton-phonon scattering with K$\Lambda$  excitons (dashed line in Fig. \ref{schema} (a)). This is due to the very efficient scattering of K-$\Lambda$ to K-K' excitons assisted by acoustic and optical $\Lambda$ phonons.
 
So far, we have discussed the exciton formation and thermalization for the exemplary TMD material WSe$_2$ and at the exemplary temperature of \unit[77]{K}. Our calculations reveal that at room temperature, the same qualitative behavior can be observed, cf. appendix. At the elevated temperature, the higher occupation of $\Gamma$ and zone edge phonons leads to several additional features: First, the region for the formation of incoherent K-K excitons is larger reflecting the broader spectral width of the coherent exciton states. Second, the relaxation to lower energetic states, driven mainly by phonon emission, becomes faster due to the $(1+n^\alpha_\mathbf{q})$ dependence of the emission rates, which increase at enhanced phonon occupations. Third, the thermalized exciton distribution becomes broader, which is consistent to expectations from a simple free Bose gas, where the width of the Bose distribution is given by the thermal energy $kT$.

Furthermore, the exciton dynamics in MoSe$_2$ turns out to be qualitatively different, cf. Fig. \ref{valleys} (b) (and the appendix for the time- and energy-dependent dynamics of exciton occupations). 
 Here, most excitons are formed in the K-K states via scattering with optical and acoustic $\Gamma$ phonons. In contrast to tungsten-based TMDs,\cite{Selig2016}  the K-K excitons are energetically lowest states in MoSe$_2$. Here, the K-$\Lambda$ excitons are located approximately $\unit[130]{meV}$ higher in energy. As a result, the steady state density of K-$\Lambda$ excitons is rather small, since these exciton states are only accessible with multiple phonon scattering events. However, we find a relatively large incoherent K-K' exciton occupation, since the K-K' excitons lie only approximately $\unit[7]{meV}$ above the K-K excitons enabling more phonon-induced scattering channels. 
In both tungsten- and molybdenum-based TMDs, the total density of created incoherent excitons is lower than the optically induced coherent exciton density. The reason lies in the radiative decay of the coherent excitons that takes part at the same time as the formation of incoherent excitons and is therefore in direct competition with this process.
At lower temperatures, the radiative recombination becomes the dominant channel drastically reducing the efficiency of the formation of incoherent excitons.

\begin{figure}[t!]
 \begin{center}
\includegraphics[width=\linewidth]{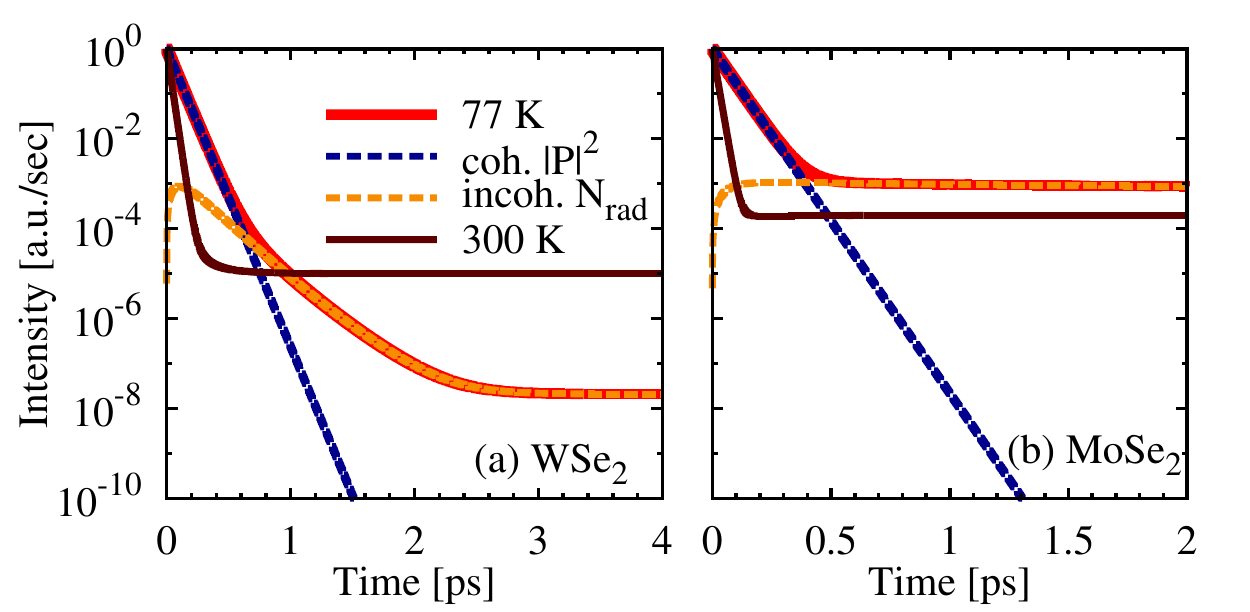}
 \end{center}
 \caption{\textbf{Time-dependent photoluminescence.} The total luminescence (solid red lines) in (a) WSe$_2$ and (b) MoSe$_2$  consists of a coherent  (dashed blue) and an incoherent part (dashed yellow). While the first dominates the PL in the first few hundreds of femtosecond, the latter takes over and is the crucial mechanism on a picosecond timescale. As a comparison, the brown curve shows the total luminescence at \unit[300]{K} illustrating the different temperature trend in  WSe$_2$ and MoSe$_2$.}
 \label{figPL}
\end{figure}

Having determined the dynamics of coherent and incoherent excitons, we can now investigate the  temporal evolution of the light emitted from TMDs. The total photoluminescence is given by  both coherent  $|P_\mathbf{0}|^2$ and  incoherent excitons $ N_{\text{rad}}=\sum_{\mathbf{Q}<K_c} N_\mathbf{Q}$ within the light cone, cf. Eq. (\ref{eqPL}). We reveal that at \unit[77]{K} and for times shorter than approximately \unit[1]{ps} the photoluminescence in WSe$_2$ is dominated by coherent light emission exceeding the incoherent contribution by orders of magnitude, cf. Fig. \ref{figPL}(a) (note the logarithmic scale of the y axis).  This is in line with the calculated exciton densities in Fig. \ref{valleys}.  The coherent part of the luminescence decays rapidly on a timescale of a few tens of femtoseconds. This excitonic coherence lifetime $T_2$ is determined by the radiative decay and phonon-induced dephasing of the excitonic polarization.\cite{Selig2016} It is reduced from approximately \unit[60]{fs} at \unit[77]{K} to \unit[30]{fs} at \unit[300]{K}. This is due to a stronger exciton-phonon scattering at higher temperatures.\cite{Selig2016}
For comparison, Fig. \ref{figPL} (b) shows the luminescence in MoSe$_2$. Here, we find a dominant coherent emission up to \unit[0.5]{ps}. The coherence lifetime is approximately \unit[60]{fs} at \unit[77]{K} and decreases to \unit[15]{fs} at room temperature. This very short lifetime can be ascribed to the very efficient formation of incoherent K-K excitons via emission and absorption of acoustic and optical $\Gamma$ phonons.\cite{Selig2016}

The incoherent part of the photoluminescence decays on a much slower timescale of a few tens of nanoseconds and corresponds to the exciton lifetime $T_1$. The reason for this slow decay is the complex interplay between radiative recombination and exciton-phonon scattering, cf. Eq. (4) in the appendix: First, phonon-driven thermalization of the incoherent exciton occupation $ N_\mathbf{Q}$ takes place and is then followed by a radiative decay of incoherent excitons within the light cone $ N_{\text{rad}}$. Since now the incoherent excitons are not thermalized any more, exciton-phonon scattering again refills the empty states within the light cone. Following this mechanism on longer timescales, this leads to an effective decay of the total exciton occupation and thus to a decay of the exciton density
within the light cone.
\begin{figure}[t!]
 \begin{center}
\includegraphics[width=\linewidth]{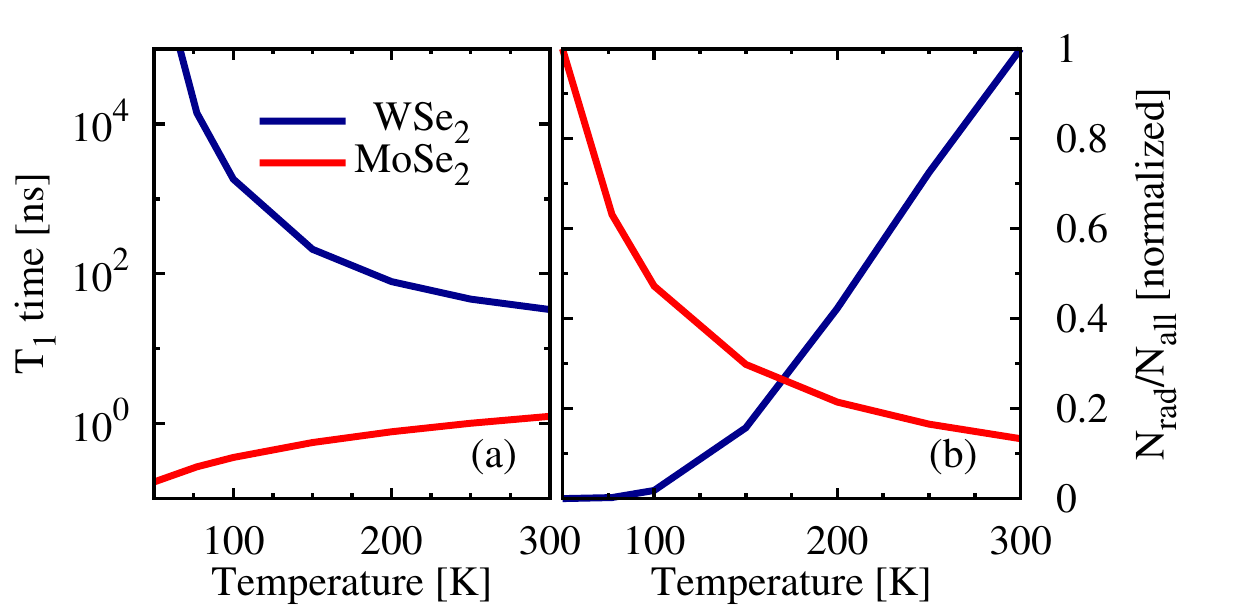}
 \end{center}
 \caption{\textbf{Luminescence yield and decay time.}  Temperature-dependent (a) decay time $T_1$ of the incoherent exciton population after thermalization for WSe$_2$ (blue) and MoSe$_2$ (red)  and (b) yield of photoluminescence defined as the ratio of bright excitons $ N_{\text{rad}}$ and all generated excitons $ N_{\text{all}}$.}
 \label{figQY}
\end{figure}

Now, we investigate the temperature dependence of the exciton lifetime $T_1$ and the photoluminescence yield  for WSe$_2$ in direct comparison with MoSe$_2$, cf. Fig. \ref{figQY}.
We find an opposite temperature behavior for the two considered TMDs: For MoSe$_2$ (WSe$_2$), the $T_1$ time increases (decreases) with temperature ranging from \unit[260]{ps} (\unit[13]{ms}) at \unit[50]{K} to \unit[1.2]{ns} (\unit[33]{ns}) at \unit[300]{K}, cf. Fig. \ref{figQY}(a). The increased $T_1$ in MoSe$_2$ can be traced back to the reduced exciton density within the light cone at elevated temperatures resulting in a less efficient radiative decay and thus a longer exciton lifetime. 
In the case of WSe$_2$, the situation is different, since the energetically lowest states are dark. Thus, higher temperatures are favorable for filling up the bright states within the light cone resulting in a faster radiative coupling and a shorter $T_1$ time. Note that we focus  here on intrinsic exciton properties in the low excitation regime neglecting non-radiative relaxation or defect-assisted exciton recombination channels. 
In many experimentally accessible situations, the relaxation of the exciton occupation will be much faster than the radiative recombination rates,\cite{AroraWSe22015,Zhang2015} and therefore the predicted intrinsic $T_1$ time has not been measured yet.

The yield of the luminescence is defined as the ratio of the time-integrated luminescence intensity to the amount of optically created excitons. Assuming comparable non-radiative recombination channels for all excitons, we estimate the luminescence yield by the ratio between the density of bright excitons $ N_{\text{rad}}$ and all generated excitons $ N_{\text{all}}$. Our calculations reveal that interestingly the yield in WSe$_2$ increases with temperature Fig. \ref{figQY} - in contrast to the behavior of conventional semiconductors. This can be directly attributed to the existence of energetically lower lying dark excitonic states in tungsten-based TMDs.\cite{Selig2016} While at low temperatures, most excitons are located at the K-$\Lambda$ states (Fig. \ref{valleys}(a)), at enhanced temperatures more and more excitons are occupying the bright states within the light cone increasing the efficiency for photoemission.  In contrast, in MoSe$_2$, where  bright states are the energetically lowest, the luminescence yield exhibits the temperature dependence expected from conventional materials, i.e.  a decreased luminescence yield at higher temperatures. Since the exciton distribution generally becomes broader at elevated temperatures,  the occupation of the bright states becomes smaller reducing the luminescence yield.
Our findings are in excellent agreement with recent experimental studies.\cite{AroraMoSe22015,AroraWSe22015,Zhang2015}
The observed opposite behavior of WSe$_2$ and MoSe$_2$ is the remarkable consequence of an extraordinary exciton landscape in TMD materials. 

In conclusion, we have presented a microscopic study revealing time- and energy-resolved formation and thermalization of bright and dark excitons in monolayer transition metal dichalcogenides. Exploiting the gained insights, we shed light on many-particle processes behind the photoluminescence in two representative TMD materials.  In particular, we demonstrate the presence of dark intervalley excitonic states that are located below optically accessible bright excitons in tungsten-based TMDs. In contrast to conventional semiconductors, these materials show an enhanced luminescence quantum yield and a reduced exciton lifetime at elevated temperatures.

\section*{Acknowledgments}
We acknowledge financial support from the Deutsche
Forschungsgemeinschaft (DFG) through SFB 787 (MS,MR, AK). This project has also received
funding from the European Unions Horizon 2020 research
and innovation programme under grant agreement No
696656 - Graphene Flagship (EM) and the Swedish Research Council (GB, EM). Finally,
MS  gratefully acknowledges inspiring discussions with Samuel Brem (Chalmers University).

\section{Appendix}

To access the exciton dynamics in TMDs, we define a many-particle Hamilton operator including: the dispersion of electrons,\cite{Kormanyos2015} phonons,\cite{Li2013,Jin2014} and photons,  the interaction of electrons with a classical electromagnetic field,\cite{Berghauser2014,Selig2016,Thranhardt2000}  the carrier-carrier interaction treated within the Hartree Fock approximation,\cite{Berghauser2014}  the carrier-phonon interaction treated in an effective deformation potential approximation,\cite{Li2013,Jin2014} and  the quantum mechanical interaction of electrons and photons.\cite{Thranhardt2000}
Exploiting the Heisenberg equation of motion and the fundamental commutation relations, we derive time- and momentum-resolved equations for the microscopic polarization $P_\mathbf{Q}$ and the 
incoherent exciton occupation
$ N_{\mathbf{Q}}$. 
We focus on the low excitation regime, where all terms in third order to the exciting
field can be neglected. In this regime, exciton-exciton and exciton-electron interactions are weak and can be neglected.\cite{Thranhardt2000}
In this limit, the Bloch equation for the microscopic polarization reads:

\begin{align}
&\partial_t P_\mathbf{Q}(t)=\frac{1}{i \hbar}\left( \frac{\hbar^2 \mathbf{Q}^2}{2M}+E - i \gamma_{\text{rad}}\,\delta_\mathbf{Q,0}\right) P_\mathbf{Q}(t) \nonumber \\
&+ \sum_\mathbf{q} \varphi^{*}_\mathbf{q} \,\mathbf{M}^{cv}_\mathbf{q} \cdot \mathbf{A}(t)  \,\delta_{\mathbf{Q},0} \nonumber\\
&-\frac{\pi}{\hbar}\sum_{\mathbf{K},\alpha,\pm} |g^{\alpha}_\mathbf{K}|^2  P_\mathbf{Q}(t) (\frac{1}{2}\pm\frac{1}{2}+n_\mathbf{K}^{\alpha}) \delta(\Delta E_\mathbf{Q+K,Q}^{\alpha \pm}).\label{CoExc}
\end{align}
The first line describes the oscillation of the excitonic polarization, where $E$ is the energy of the investigated $A_{1s}$ exciton and $\gamma_{\text{rad}}$ the radiative decay rate of the polarization. The latter is calculated by self-consistent solution of the Maxwell equations and the Bloch equation.\cite{Knorr1996,Selig2016} The second line describes the excitation of the polarization with a classical external field that is determined by the scalar product of the optical matrix element $\mathbf{M}^{cv}_\mathbf{q}=\frac{e_0}{m_0} \langle \mathbf{q} c | \nabla | \mathbf{q} v  \rangle $  and the vector potential $\mathbf{A}(t)$ of the exciting field.
Finally, the third line describes the decay of the microscopic polarization due to exciton-phonon scattering leading to the formation of incoherent exciton densities. Here,  $g^{\alpha}_\mathbf{q}$ denotes the exciton-phonon coupling element,  $n_\mathbf{q}^{\alpha}$ the phonon occupation with the momentum $\mathbf{q}$ and the mode $\alpha$, and finally  $\Delta E_\mathbf{K_1,K_2}^{\alpha \pm}=E_\mathbf{K_1}-E_\mathbf{K_2}\pm \hbar \Omega^{\alpha}_\mathbf{K_1-K_2} $ expresses the condition for the energy conservation, where $\hbar \Omega^{\alpha}_\mathbf{q}$ is the energy of the involved phonon, cf. Ref. \onlinecite{Selig2016} for more details on the implementation of exciton-phonon scattering. 

In the same limit of low excitation, we derive the equation of motion for the incoherent exciton densities yielding:

\begin{align}
&\partial_t  N_{\mathbf{Q}}(t)=\nonumber \\
&\frac{2\pi}{\hbar}\sum_{\mathbf{K},\alpha,\pm}|g^{\alpha}_\mathbf{K}|^2 |P_{\mathbf{Q+K}}(t)|^2  \left(\frac{1}{2}\pm\frac{1}{2}+n_\mathbf{K}^{\alpha}\right) \delta\left(\Delta E_\mathbf{Q,Q+K}^{\alpha \pm}\right)\nonumber \\
&-\frac{2 \pi}{\hbar}\sum_{\mathbf{K},\alpha,\pm}|g^{\alpha}_\mathbf{K}|^2  N_\mathbf{Q}(t)\, \left(\frac{1}{2}\pm\frac{1}{2}+n_\mathbf{K}^{\alpha}\right)  \delta\left( \Delta E_\mathbf{Q+K,Q}^{\alpha \pm}\right)\nonumber \\
&+\frac{2 \pi}{\hbar}\sum_{\mathbf{K},\alpha,\pm}|g^{\alpha}_\mathbf{K}|^2  N_\mathbf{Q+K}(t)\, \left(\frac{1}{2}\pm\frac{1}{2}+n_\mathbf{K}^{\alpha}\right)   \delta \left( \Delta E_\mathbf{Q,Q+K}^{\alpha \pm}\right)\nonumber \\
&-\frac{2\pi}{\hbar} \sum_{\mathbf{k},\sigma} |M^{\sigma}_\mathbf{Q,k}|^2\, \delta_{\mathbf{Q,k_{\parallel}}} \, N_\mathbf{Q}(t)\,  \delta(E_\mathbf{k_\parallel}-\hbar \omega_\mathbf{k}^{\sigma}).\label{InExc}
\end{align} 
The first line describes the formation of incoherent exciton densities due to the non radiative decay of coherent excitons. The second and the third line contain exciton phonon scattering terms where the second contributes to out scattering and the third to in scattering processes. In all three lines the $+$ terms denote phonon emission processes, whereas the $-$ terms denote phonon absorption processes. The last line describes the radiative decay of incoherent excitons due to spontaneous emission of photons.

\begin{figure}[t!]
  \begin{center}
     \includegraphics[width=\linewidth]{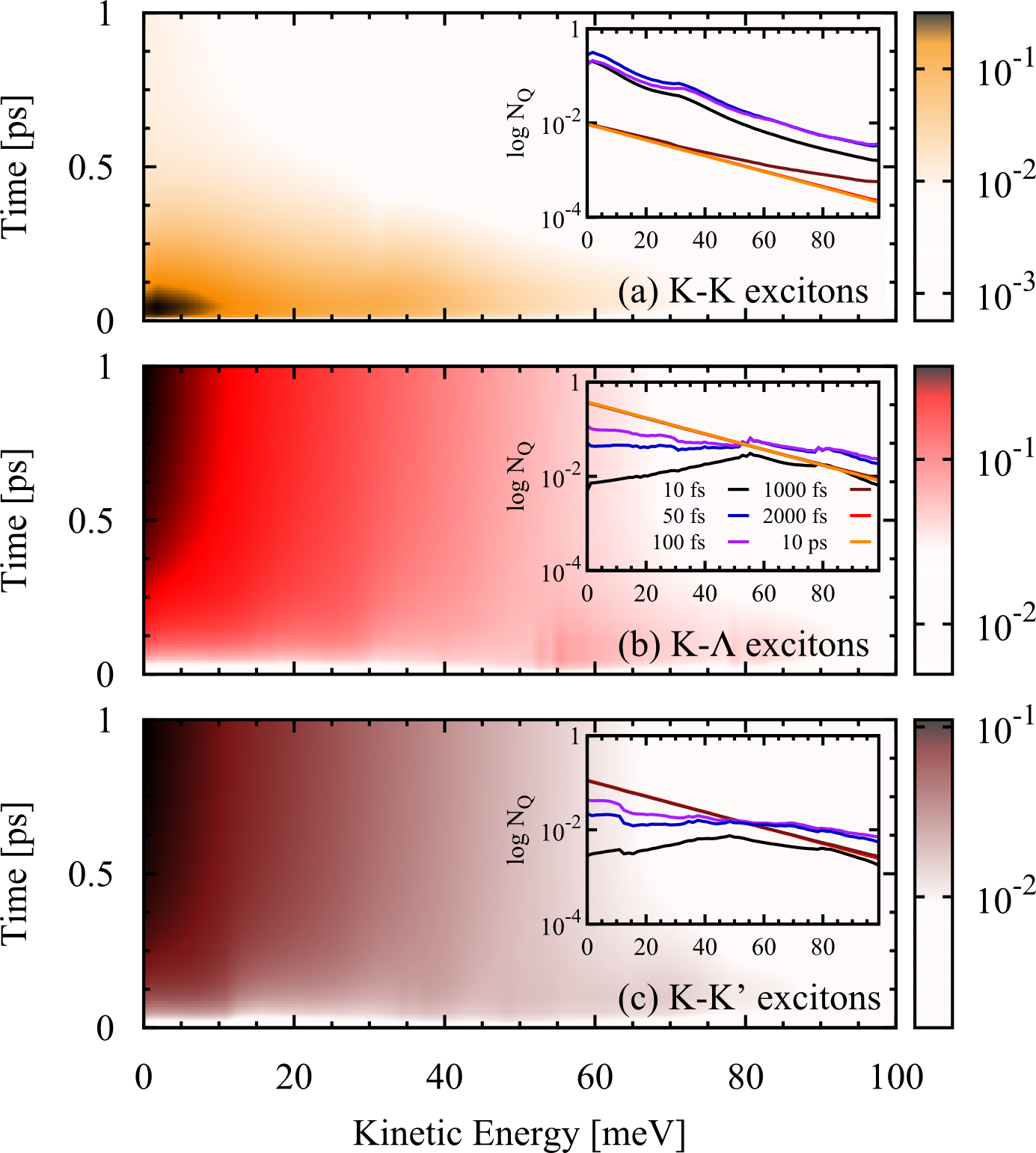}
   \end{center}    
  \caption{\textbf{Exciton dynamics in WSe$_2$ at \unit[300]{K}}}
 \label{supp_Con300}
\end{figure}

Now, we discuss the exciton formation and thermalization in WSe$_2$ at \unit[300]{K}, cf. Fig. \ref{supp_Con300}. Phonon-mediated exciton formation takes place at the same energetic spectral positions as at \unit[77]{K} (discussed in the main text), since the excitonic bandstructure and the phonon dispersion do not depend on temperature. Higher temperatures result in larger phonon occupations and thus a more efficient exciton-phonon scattering can take place resulting in a faster exciton formation and thermalization. Here, we observe already \unit[1]{ps} after the optical excitation a Bose-Einstein distribution of the exciton occupations.

\begin{figure}[t!]
  \begin{center}
     \includegraphics[width=\linewidth]{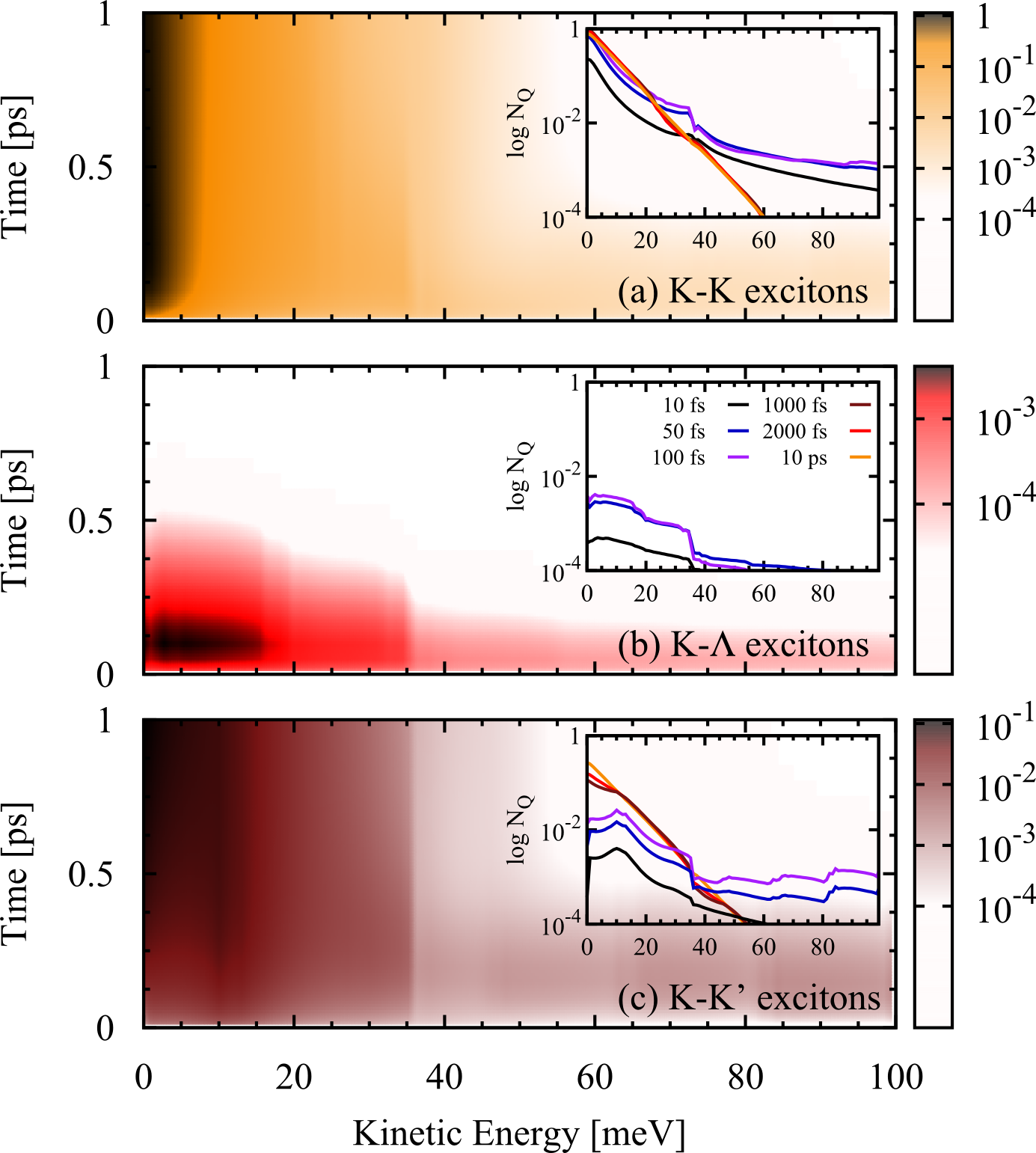}
   \end{center}    
  \caption{\textbf{Exciton dynamics in MoSe$_2$} at \unit[77]{K}.}
 \label{supp_Con77Mo}
\end{figure}

\begin{figure}[t!]
  \begin{center}
     \includegraphics[width=\linewidth]{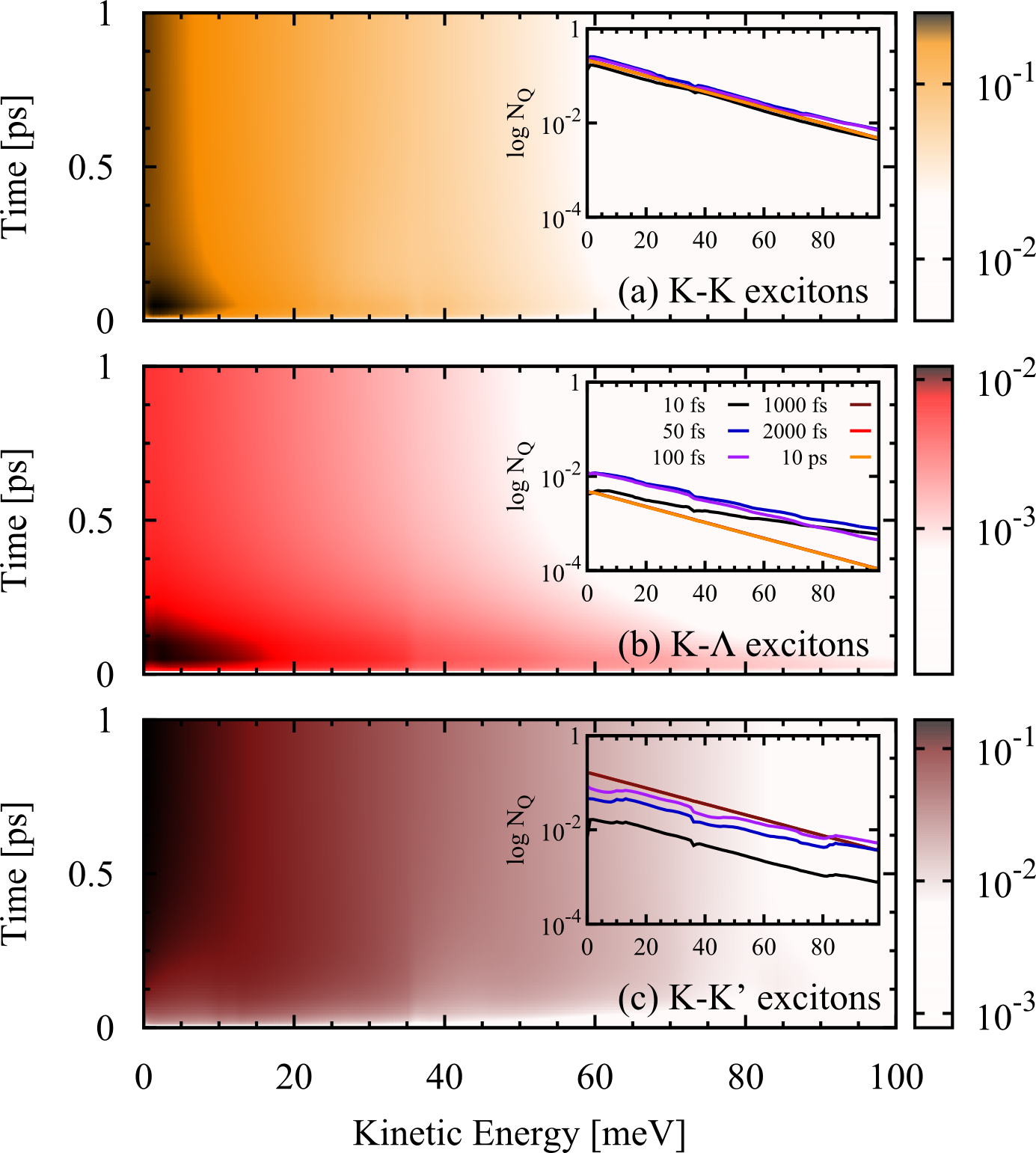}
   \end{center}    
  \caption{\textbf{Exciton dynamics in MoSe$_2$} at \unit[300]{K}.}
 \label{supp_Con300Mo}
\end{figure}

\begin{figure}[t!]
  \begin{center}
     \includegraphics[width=\linewidth]{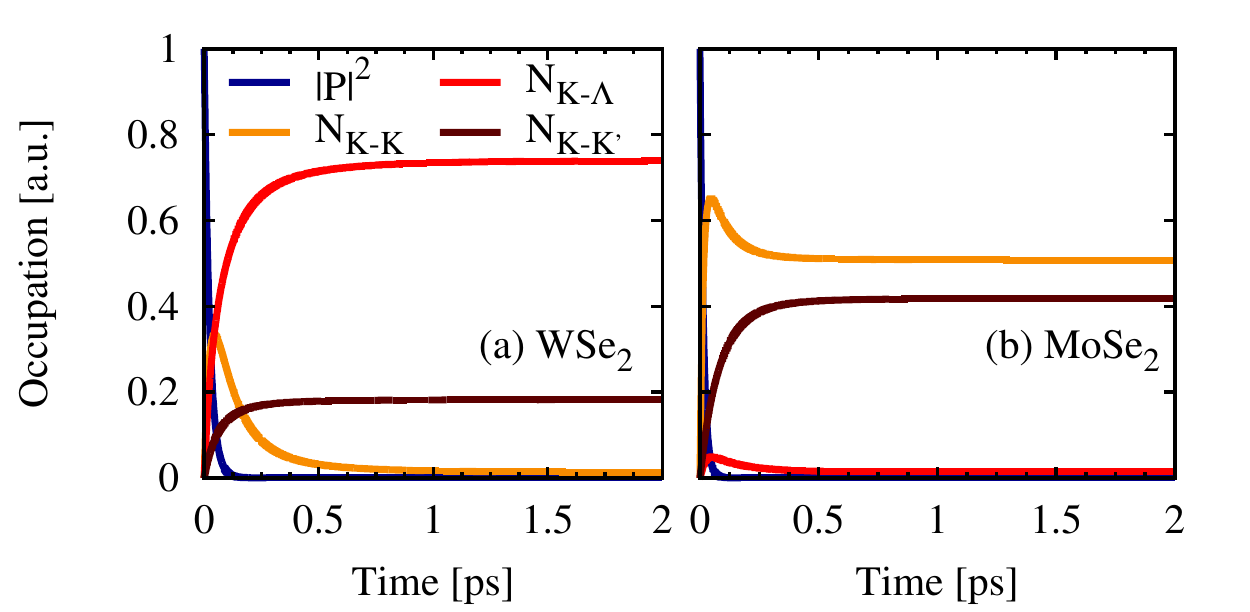}
   \end{center}    
  \caption{\textbf{Exciton densities in different valleys in WSe$_2$ and MoSe$_2$ at \unit[300]{K}}}
 \label{supp_val300}
\end{figure}

In MoSe$_2$, the exciton formation occurs mainly at the excitonic $\Gamma$ valley resultig in $K-K$ excitons. This is due to the efficient scattering with intravalley acoustic phonons, cf. Figs \ref{supp_Con77Mo} and \ref{supp_Con300Mo}. Increased phonon-assisted exciton formation occurs around \unit[36]{meV}, which coincides with the optical phonon energy in MoSe$_2$ \cite{Jin2014}. Furthermore, we find exciton formation in K-K' states at \unit[10]{meV} above the minimal excitonic energy through the absorption of acoustic $K$ phonons. The energy of the latter is \unit[16]{meV} and the separation between the K-K and the K-K' states is about \unit[7]{meV}. The small amount of created K-$\Lambda$ excitons can be traced back to off-resonant absorption processes. Scattering with phonons leads to thermalization of the exciton occupation, which reaches a steady state after \unit[2]{ps}. Here, K-K states show the highest occupation, since they are the energetically lowest lying states in MoSe$_2$, cf. Fig. \ref{supp_val300}.
At \unit[300]{K}, the formation of excitons is already finished after some tens of fs due to a more efficient exciton-phonon scattering resulting from a larger phonon occupations at higher temperatures. The thermalization of the exciton occupations is also faster resulting in thermalized steady state  exciton densities already after approximately \unit[500]{fs}, cf. Fig. \ref{supp_val300}.

Note that we treat the exciton-phonon interaction in second-order Born-Markov approximation \cite{Kochbuch} that has been already demonstrated to be an excellent approximation to reproduce the experimentally measured homogeneous linewidths of excitonic resonances.\cite{Selig2016} At higher temperatures, multiple exciton-phonon processes might become important, which would further accellerate the formation and thermalization of excitons.

\bibliographystyle{unsrt}


\end{document}